\documentclass[11pt]{article}
\usepackage{authblk}
\usepackage{graphicx}
\usepackage{epstopdf}

\begin{document}

\title{Comment on \textquotedblleft Classification of Cosmic Scale Factor
via Noether Gauge Symmetries\textquotedblleft\ [Int. J. Theor. Phys. 54,
2343 (2015)]}
\author[1,2]{A Paliathanasis\thanks{%
paliathanasis@na.infn.it}}
\author[3]{K Krishnakumar\thanks{%
krishapril09@gmail.com}}
\author[4,5,6]{ PGL Leach\thanks{%
leach@ucy.ac.cy}}

\affil[1]{Dipartimento di Fisica, Universita' di Napoli,\textquotedblleft
Federico II\textquotedblright, Complesso Universitario di Monte S. Angelo, Via Cinthia 9,
I-80126, Napoli, Italy}
\affil[2]{Istituto Nazionale di Fisica Nucleare
(INFN) Sez. di Napoli, Complesso Universitario di Monte S. Angelo, Via
Cinthia 9, I-80126, Napoli, Italy}
\affil[3]{Department of Mathematics, Pondicherry University, Kalapet Puducherry 605 014, India}
\affil[4]{Department of Mathematics and
Institute of Systems Science, Durban
University of Technology, PO Box 1334, Durban 4000, Republic of South Africa}
\affil[5]{School of Mathematics, Statistics and Computer Science, University
of KwaZulu-Natal, Private Bag X54001, Durban 4000, Republic of South Africa}
\affil[6]{Department of Mathematics and Statistics, University of Cyprus,
Lefkosia 1678, Cyprus}

\renewcommand\Authands{ and }
\maketitle

\begin{abstract}
We discuss the relationship between the Noether point symmetries of the
geodesic Lagrangian, in a (pseudo)Riemannian manifold, with the elements of
the Homothetic algebra of the space. We observe that the classification
problem of the Noether symmetries for the geodesic Lagrangian is equivalent
with the classification of the Homothetic algebra of the space, which in the
case of a Friedmann-Lema\^{\i}tre-Robertson-Walker spacetime is a well-known
result in the literature.
\end{abstract}

\bigskip
Recently the classification of the Noether symmetries for the
geodesic Lagrangian in Friedmann-Lema\^{\i}tre-Robertson-Walker (FRW)
spacetimes was performed in \cite{ref1}.

FRW spacetime describes an isotropic universe for which the fundamental line
element is
\begin{equation}
ds^{2}=-dt^{2}+\frac{a^{2}\left( t\right) }{\left( 1+ \mbox{$\frac{1}{4}$}K%
\mathbf{x}^{2}\right) ^{2}}\left( dx^{2}+dy^{2}+dz^{2}\right) ,
\label{eq.01}
\end{equation}%
where $a\left( t\right) $ is the scale factor of the universe, $K$ is the
spatial curvature and $\mathbf{x}^{2}=x^{2}+y^{2}+z^{2}$. The spacetime with
line element (\ref{eq.01}) admits a six-dimensional Killing algebra, for
arbitrary $a\left( t\right) $. It is the Killing algebra of the
three-dimensional Euclidian space for $K=0$, or the Killing algebra of the
three-dimensional space of constant curvature, $K,$ with $K\neq 0$. The
classification of the Killing vectors (KVs), the Homothetic vector (HV) and
the Conformal Killing vectors (CKVs) can be found in \cite{sunil}.

The geodesic Lagrangian for a (pseudo)Riemannian manifold is defined as%
\begin{equation}
L\left( x^{i},\dot{x}^{i}\right) = \frac{1}{2}g_{ij}\left( x^{k}\right) \dot{%
x}^{i}\dot{x}^{j}  \label{eq.02}
\end{equation}%
in which $i,j,k=1,2,...,\dim g_{ij}$ and overdot means total derivative with
respect to the affine parameter \textquotedblleft $s$\textquotedblright $.$
The Euler-Lagrange equations which follow from (\ref{eq.02}) are%
\begin{equation}
\ddot{x}^{i}+\Gamma _{jk}^{i}\dot{x}^{i}\dot{x}^{j}=0.  \label{eq.03}
\end{equation}%
This system describes the free-fall of a particle in a space with metric $%
g_{ij}$, i.e. the geodesic equations.

Consider the infinitesimal one-parameter point transformation in the space
of the variables $\left\{ s,x^{i}\right\} $%
\begin{equation}
\bar{s}=s+\varepsilon \xi \left( s,x^{k}\right)  \label{eq.04}
\end{equation}%
\begin{equation}
\bar{x}^{i}=x^{i}+\varepsilon \eta ^{i}\left( s,x^{k}\right)  \label{eq.05}
\end{equation}%
where the generator of the infinitesimal transformation is defined by%
\begin{equation}
X=\frac{\partial \bar{s}}{\partial \varepsilon }\partial _{s}+\frac{\partial
\bar{x}^{i}}{\partial \varepsilon }\partial _{i},  \label{eq.06}
\end{equation}%
that is,%
\begin{equation}
X=\xi \left( s,x^{k}\right) \partial _{s}+\eta ^{i}\left( s,x^{k}\right)
\partial _{i}.  \label{eq.07}
\end{equation}

By definition the generator, $X$, of the one-parameter point transformation,
(\ref{eq.04})-(\ref{eq.05}), of the Action Integral of a Lagrangian $%
L=L\left( s,x^{k},\dot{x}^{k}\right) ,~$which transforms the Action Integral
in such a way that the Euler-Lagrange equations are invariant, is called a
Noether symmetry \cite{noe}.

Mathematically, $X$ is a Noether symmetry if there exists a function $f^{i}$
such that \cite{noe}
\begin{equation}
X^{\left[ 1\right] }\left( L\right) +L\dot{\xi}=\dot{f}.  \label{eq.08}
\end{equation}%
The corresponding conservation law is
\begin{equation}
\Phi =\xi \left( \dot{x}^{k}\frac{\partial L}{\partial \dot{x}^{k}}-L\right)
-\eta ^{k}\frac{\partial L}{\partial u_{k}}+f,  \label{eq.09}
\end{equation}%
where $X^{\left[ 1\right] }$ is the first prologation/extension of $X$ in
the space of variables $\left\{ s,x^{i},\dot{x}^{i}\right\} $.

As far as concerns the Noether (point) symmetries of the geodesic Lagrangian
(\ref{eq.02}), in \cite{and1} it has been shown that there exists a unique
connection between the Noether symmetries and the Homothetic algebra of the
underlying space. Specifically, the Noether symmetries for the Lagrangian (%
\ref{eq.02}) are the vector fields%
\begin{equation}
X_{1}=\partial _{s}~,~X_{KV}=K_{I}~,~X_{HV}=2s\partial _{s}+H,  \label{eq.10}
\end{equation}

\begin{equation}
X_{GKV}=sS^{,i}\partial _{i}~,~X_{GHV}=t^{2}\partial _{s}+s\Omega
^{,i}\partial _{i},  \label{eq.11}
\end{equation}%
in which the $K^{I}$ form the Killing algebra of the space $g_{ij}$, $H$ is
a proper HV of $g_{ij}$, $S^{,i}$ is gradient KV and $\Omega ^{,i}$ is a
gradient HV. Recall that the space $g_{ij}$ can a maximum of one HV and the $%
S_{,I}^{i}$ are included in the Killing algebra of $g_{ij}$. Finally, the
vector field $X_{1}$ is the autonomous symmetry and the corresponding
conservation law is the Hamiltonian function.

Therefore the problem of the classification of the Noether symmetries for
the geodesic Lagrangian (\ref{eq.02}) is equivalent with the classification
problem of the KVs and HV for the metric $g_{ij}$.

For the line element (\ref{eq.01}) the geodesic Lagrangian is%
\begin{equation}
L=-\frac{1}{2}\dot{t}^{2}+\frac{a^{2}\left( t\right) }{\left( 1+%
\mbox{$\frac{1}{4}$}K\mathbf{x}^{2}\right) ^{2}}\left( \dot{x}^{2}+\dot{y}%
^{2}\dot{+}z^{2}\right) .  \label{eq.12}
\end{equation}%
and without loss of generality we can perform a coordinate transformation, $%
t\rightarrow \tau ,$ whereby (\ref{eq.12}) takes the simpler form%
\begin{equation}
L=R^{2}\left( \tau \right) \left[ -\frac{1}{2}\dot{\tau}^{2}+\frac{1}{\left(
1+\mbox{$\frac{1}{4}$}\mathbf{x}^{2}\right) ^{2}}\left( \dot{x}^{2}+\dot{y}%
^{2}\dot{+}z^{2}\right) \right] .  \label{eq.13}
\end{equation}

The Lie and Noether symmetries for the geodesic Lagrangian, (\ref{eq.13}),
have been given in \cite{and1} as an illustrative example for the main
results of the paper. The classification is based upon \cite{sunil}. Hence,
under the inverse transformation $\tau \rightarrow t,$ the results of \cite%
{and1} can be used for the Lagrangian, (\ref{eq.12}).

Furthermore, in \cite{ref1}, the authors claim that the Lagrangian, (\ref%
{eq.12}), admits only two Noether symmetries for an arbitrary scale factor, $%
a\left( t\right) $. However, this is not true because, as we discussed
above, the FRW spacetime (\ref{eq.01}) always admits a six-dimensional
Killing algebra. Hence, for an arbitrary functional form of $a\left(
t\right) $, the Lagrangian (\ref{eq.12}) admits seven Noether point
symmetries.

Finally we note that the classification of the Lie symmetries for the
geodesic equations (\ref{eq.03}) in a FRW spacetime (\ref{eq.01}) can be
found in \cite{and1} and it is equivalent with the classification problem of
the special projective algebra of the space \cite{mar1,proj1}.

\noindent\textbf{Remark:} In \cite{ref1} the Noether symmetries are termed
`gauge' symmetries. This is incorrect terminology as the function, $f$, of
Noether's Theorem is not a gauge function. It is a boundary term introduced
to allow for the infinitesimal changes in the value of the Action Integral
produced by the infinitesimal change in the boundary of the domain caused by
the infinitesimal transformation of the variables in the Action Integral.
This is clearly stated in Noether's paper of nearly 100 years ago. Noether's
Theorem really comes in two parts. The first is the equation for the
calculation of the symmetries and the second is the formula for a
conservation law. The second invokes the Euler-Lagrange equation but not the
first. A further remark is that Noether allowed for generalised symmetries
and not just point symmetries. It is a pity that writers on a subject
connected with Noether's Theorem do not read the original paper.

\subsubsection*{Acknowledgements}

AP and KK acknowledge Prof. PGL\ Leach, Sivie Govinder, as also DUT for the
hospitality provided and the UKNZ of South Africa for financial support. The
research of AP is financial supported by INFN. KK thanks the University
Grants Commission for providing a UGC-Basic Scientific Research Fellowship
to perform this research work.


\begin{thebibliography}{9}
\bibitem{ref1} A. Jhangeer, M.F. Shamir, T. Naz and N. Iftikhar, Int. J.
Theor. Phys \textbf{54} 2343 (2015)

\bibitem{sunil} R. Maartens and S.D. Maharaj, Class. Quant. Grav. \textbf{3}%
, 1005 (1986)

\bibitem{noe} E. Noether, Nachr. v.d. Ges. d. Wiss. zu Gottingen \textbf{235}%
, (1918)

\bibitem{and1} M. Tsamparlis and A. Paliathanasis, Gen. Relativ. Gravit.
\textbf{42}, 2957 (2010)

\bibitem{mar1} R. Maartens, J. Math. Phys. 28, 2051 (1987)

\bibitem{proj1} G.S. Hall, Class. Quant. Grav. \textbf{17}, 4637 (2000)
\end{thebibliography}
\end{document}